\documentstyle{article}
\begin{document}
\openup6pt 

\title {Strong field gravitational lensing in Scalar-tensor theories} 
\author{Kabita Sarkar \footnote{Department of Mathematics}
and Arunava Bhadra\footnote{Administrative Block, Electronic address: {\em
aru\_bhadra@yahoo.com}}
\\
University of North Bengal, Siliguri 734430 INDIA \\
} 
\date{}
\maketitle

\begin{abstract}
Strong field gravitational lensing in the Brans-Dicke scalar tensor theory has been studied. The deflection angle for photons passing very close to the photon sphere is estimated for the static spherically symmetric space-time of the theory and the position and magnification of the relativistic images are obtained. Modeling the super massive central object of the galaxy by the Brans-Dicke space-time, numerical values of different strong lensing observable are estimated. It is found that against the expectation there is no significant scalar field effect in the strong field observable lensing parameters. This result raises question on the potentiality of the strong field lensing to discriminate different gravitational theories.
\end{abstract}

Key Words: Gravitational lensing, Strong field, Scalar-tensor theory  \\ 
PACS numbers: 98.62.Sb, 95.30.Sf, 04.50.+h \\

\section*{I. Introduction} 
Scalar tensor (ST) theories of gravitation [1], in which gravity is mediated by one or several long range scalar field(s) in addition to the usual tensor fields present in Einstein's theory, are widely considered as most viable alternatives to Einstein's general theory of relativity (GR). The inclusion of scalar fields in the gravitational sector is justified from the fact that their presence is inevitable in most of the theoretical attempts to unify gravity with other fundamental interactions, such as the superstring theory, supergravity or modern revival of the Kaluza-Klein theory. Cosmological observations too insist for the introduction of long range scalar field; all most all scenarios of cosmological inflation are based on scalar field.\\
The introduction of scalar fields obviously leads corrections to general relativistic dynamics. These deviations from GR can be expressed in terms of coupling function $\omega(\varphi)$ that characterizes a ST theory and represents the strength of the coupling between the scalar field ($\varphi$) and the curvature. The experimental observations, however, suggest that the contribution of scalar field is not more than a very small fraction of that of the tensor field, if not zero, and as a result ST theories are severely constrained by the requirement that $\mid\omega(\varphi)\mid$ is very large. The VLBI observations of radio wave deflection demand $\vert \omega \vert > 500$ [2] whereas the recent conjunction experiment with Cassini spacecraft [3] imposes even harder restriction $\vert \omega \vert > 5 \times 10^4$. Such a limit on $\omega(\varphi)$ obviously raises doubt on the existence of gravitational scalar field because in the limit $ \vert \omega \vert \rightarrow \infty$ post-Newtonian expansions of ST gravity reduces to those of GR [1] (however, sea also [4]). Small contribution of scalar field (relative to that of tensor field) has been explained through the idea [5] that most of the ST theories are cosmologically evolved toward a state with practically no scalar admixture to gravity during matter dominated era. This means that for a large class of ST theories $\omega(\varphi)$ cosmologically evolves toward a very large value. So present lower bound on $\omega(\varphi)$ does not rule out ST gravity. \\
The theoretical speculation [5] is that at the present epoch $\omega(\varphi) \sim 1.4 \times 10^{6} (\Omega^{3}_{o}/H_{o})^{1/2}$ where $\Omega_{o}$ is the ratio of the current density to the closure density and $H_{o}$ is the Hubble constant in units of $100 \; km \; s^{-1}Mpc^{-1}$. So in the present time $\omega(\varphi)$ could have finite large value. For such large $\omega(\varphi)$, deviations of gravity in the solar system from the general relativistic values are extremely small. This motivates new searches [6] for small deviations at levels better than $10^{-5}$ or even $10^{-7}$ of the post Newtonian effects but obviously it is a very difficult task. However, even for large $\omega(\varphi)$, ST theories may produce interesting departures from GR at the strong field scenario. Here strong field is distinguished from the weak gravity through the quantity $\frac{GM}{rc^{2}}$ [8] (in the strong field regime higher order terms in $\frac{GM}{rc^{2}}$ can not be ignored). For instance in the case of  generation of gravitational waves, ST gravity allows binary systems (consisting two massive compact objects) to emit dipole radiation whereas GR admits only quadrupole and higher angular modes. The total gravitational energy radiated by a given source is also different in these theories. As a result experiments like the Laser Interferometric Gravitational-Wave Observatory (LIGO) may discriminate these two theories or at least yield stronger bound on $\omega(\varphi)$ than are achievable from the solar system measurements [7]. \\
Gravitational lensing by a massive compact lens is considered another potential tool for studying strong fields. Theoretical investigations [9-12] suggest that while propagating close to massive compact object (e.g. a black hole) light rays take several turns around the lens before reaching the observer and as a result apart from primary and secondary images a set of infinite images on both side of the optic axis will be produced which are termed as relativistic images. Though in such situation the primary and secondary lensed images carry important information on various orders of post-post-Newtonian effects [13] but these relativistic images are the main signature of the strong field lensing. However, unless the source is almost perfectly aligned with the lens and the observer, relativistic images will be very faint as a result of high demagnification. With the indication that the radio-source Sgr A* in the galactic centre hosts a supermassive object (black hole) of $3.6 \times 10^{6}$ solar masses [14] a possibility has developed of studying lensing phenomena in the strong gravity regime. It is thus imperative to investigate the effects of scalar field in strong field situation and to look for its possible observational signatures.  This is precisely the aim of the present work. \\ 
An essential pre-requisite for studying strong field lensing is to have knowledge of exact explicit solution(s) of the theory. But no such solution is currently available for the generalized ST gravity. Hence our discussion would be restricted only to the simplest version of ST gravity that developed by Jordan, Fierz, Brans and Dicke, and is commonly known as Brans-Dicke (BD) theory [15] (this is a standard approach, see for example [7]). In the BD theory, there is only one scalar field and $\omega(\varphi)$ is assumed to be a fixed constant.  It is to be noted that in dealing with the scalar-tensor theories in general and the BD theory in particular, one envisages two types of frames, viz., the Jordan and Einstein frames, which are conformally connected. Sometimes it is mathematically more preferable to use the Einstein representation as the spin-2 and spin-0 fields are decoupled in the later frame from each other and the behavior of the fields are more readily manageable but experimentally observed quantities are those that are written in the Jordan frame [16, 5] which is also known as a physical frame. Ordinary (normal) matter has universal coupling to (physical) metric in the Jordan frame which implies that test particles follow geodesics of the geometry and physical rods and clocks measure the Jordan frame metric. The weak equivalence principle, conservation laws and the constancy of the non-gravitational constants are only preserved in this frame. An undesirable feature of the frame is that the energy density of scalar fields is not positive-definite here though there are some ambiguities in the definition of energy density itself [17]. But more importantly it has been shown very recently [18] that the Minkowski space is stable in this frame with respect to inhomogeneous scalar and tensor perturbations, at least at the linear order. Einstein frame formulation leads to a well-defined energy momentum tensor for the scalar field but in this frame ordinary matter is non-minimally coupled with the scalar field and consequently test particles do not move on geodesics of the Einstein frame metric. Virbhadra and Ellis [19] have already studied gravitational lensing numerically in strong field regime in the BD theory but in the Einstein frame. Bozza [20] too obtained analytical expressions of strong field lensing in the Einstein frame BD theory. For obvious reasons here we would like to study  strong field lensing for the BD theory in the physical frame. Since the two frames are conformally coupled, in many cases we would exploit the Einstein frame results [20] without going in to all detailed calculations starting from {\it ab initio}.  \\
The paper is organized as follows. The BD theory in Jordan and Einstein frames and its spherically symmetric vacuum solutions will be revisited in Sec. II. In Sec.III after reviewing strong field lensing technique for a general static spherically symmetric spacetime, the deflection angle for the physical metric of the Jordan frame BD theory will be obtained. Expressing the gravitational field due to super massive central object of the galaxy by the BD theory, an estimation of observational strong lensing parameters will be given in Sec.IV along with the similar estimation when the lens is represented by a Schwarzschild black hole. A discussion of the results will be made in Sec. V. 

\section*{II. The BD theory in Jordan and Einstein frames}
As mentioned in the previous section, the BD theory can be formulated in two distinguished conformal frames: the Jordan frame and the Einstein frame.  
\subsection*{IIA. The BD theory in Jordan frame}
In the Jordan conformal frame, the BD action takes the form (we use geometrized units such that $G=c=1$ and follow the signature -,+,+,+)
\begin{equation}
{\cal A}= \frac{1}{16 \pi }\int d^{4}x \sqrt{-g}\left(\varphi R+\frac{\omega }{\varphi } 
g^{\mu\nu} \varphi_{,\mu} \varphi_{,\nu}  \right) + {\cal A}_{matter}[\psi_{m}, g_{\mu\nu}]
\end{equation}
The last term is the action of the ordinary matter fields, $\psi_{m}$, which couple only to the metric $g_{\mu\nu}$ and not to the scalar field. Variation of (1) with respect to $ g^ {\mu \nu} $ and $\varphi$ gives, respectively, the field equations
\begin{equation} 
R_{\mu\nu} -\frac{1}{2}g_{\mu\nu}R= -\frac{8 \pi} {\varphi } T_{\mu \nu}
-\frac {\omega}{\varphi ^{2}}\left( \varphi_{,\mu} \varphi_{,\nu}- \frac{1}{2} g_{\mu\nu} 
\varphi^{,\sigma} \varphi_{,\sigma} \right) - 
\frac{1}{\varphi} \left( \nabla_{\mu}\nabla_{\nu}\varphi-g_{\mu \nu} \Box \varphi \right),  
\end{equation}
\begin{equation}
\Box \varphi = \frac {8\pi T}{(2\omega + 3)} 
\end{equation}
where $R$ is the Ricci scalar, $T_{\mu \nu}= \frac{2}{\sqrt{-g}}\frac{ \delta {\cal A}_{matter} }{\delta g^{\mu\nu}}$ is the physical frame energy-momentum tensor and $T$=$T_{\mu}^{\mu}$ is the trace of the matter energy momentum tensor. As evident from the above field equations that in the Jordan BD theory the scalar field acts as the source of the (local) gravitational coupling with $G \sim \varphi^{-1}$ and consequently the gravitational {\lq constant \rq} is not in fact a constant. \\
Since Birkhoff's theorem does not hold in the presence of a scalar field, several static solutions of the BD theory seems possible even in spherically symmetric vacuum situations. Four forms of static spherically symmetric vacuum solution of the BD theory are available in the literature, which are named after Brans [21] (in fact Brans class I solution was discovered jointly by Brans and Dicke [15] and hereafter we shall call it as the BD class I solution). However, recent studies [22, 23] suggest that only two classes of solution are really independent; choice of imaginary parameters in the BD class I solution leads to the class II solution whereas under a redefinition of the radial variable class III solution maps to class IV. Further by matching exterior and interior (due to physically reasonable spherically symmetric matter source) scalar fields it has been found that only the BD class I solution with certain restriction on solution parameters may represent exterior metric for a nonsingular massive object. The BD class I solution (in isotropic coordinates) is given by 
\begin{equation}
ds^{2}= -\left( \frac{1-B/\rho}{1+B/\rho} \right)^{\frac{2}{\lambda}} dt^{2} + \left( 1 + \frac{B}{\rho}\right) ^{4} \left( \frac{1-B/\rho}{1+B/\rho} \right)
^{\frac{2(\lambda -C -1)}{\lambda}} \left( d\rho^{2} +\rho^{2} d\theta ^{2} +\rho^{2} sin^{2} \theta d\phi ^{2} \right)
\end{equation}
\begin{equation}
\varphi = \varphi_{0} \left( \frac{1-B/\rho}{1+B/\rho} \right)^{\frac{C}{\lambda}}
\end{equation}
with the constraint condition
\begin{equation}
\lambda^2 =  (C+1)^{2} - C(1-\frac{\omega C}{2})
\end{equation}
where $ B$ and $ C $ are arbitrary constants. \\
Matching of exterior and interior scalar fields demands
\begin{equation}
C=-\frac{1}{\omega +2}; \; 2B/\lambda=M \; \mbox{and} \; \lambda=\sqrt{ \frac{2\omega +3}{2 \omega +4}}
\end{equation}
An important point to note is that though the BD class I solution is not the unique solution of the BD theory but it is the most general physically acceptable static spherically symmetric solution of the theory [22]. In the limit $\omega$ tends to $\infty$ this solution reduces to the Schwarzschild metric with constant scalar field. Other  claimed {\it new} spherically symmetric static vacuum  solutions of the BD theory are found essentially limiting cases of the BD class I solution [24]. \\
In general, the BD class I solution exhibits naked singularity; all curvature invariants diverge at the horizon $\rho=B$ (it exhibits black hole nature only when $ -2 > \omega > -(2+\frac{1}{\sqrt{3}})$ [22], such small values of $ \omega $ is already ruled out by observations). Here it is worthwhile to mention that though the naked singularity is undesirable to many physicists but whether a naked singularity occurs generically in a physically realistic collapse is a subject of considerable debate [25]. Since no proof of cosmic censorship hypothesis is avalable, only observation can give a final verdict on the issue.  The BD class I solution with coupling constant $\omega$ less than $-1.5$ (excluding the point $\omega =2$) also gives rise to physically viable traversable wormhole geometry though it is not very suitable for interstellar travel [26].\\
Under the coordinate transformation 
\begin{equation}
r=\rho \left( 1 + \frac{B}{\rho} \right) ^{2}
\end{equation}
the BD class I metric takes the form
\begin{equation}
ds^{2}= -\left( 1-\frac{4B}{r} \right)^{\frac{1}{\lambda}} dt^{2} + \left( 1 - \frac{4B}{r}\right) ^{-\frac{C+1}{\lambda}} dr^{2} + \left( 1-\frac{4B}{r}\right)
^{1-\frac{C+1}{\lambda}} \left( r^{2} d\theta ^{2} +r^{2} sin^{2} \theta d\phi ^{2} \right)
\end{equation}
which is mathematically more convenient for studying the strong field lensing.

\subsection*{IIB. The BD theory in Einstein frame}
Defining a conformally related metric through what has been known as Dicke transformations 
\begin{equation}
\tilde{g}_{\mu \nu}=\varphi g_{\mu \nu}
\end{equation}
and a redefinition of the scalar field
\begin{equation}
d\tilde{\varphi}=\left[\frac {2\omega+3}{2}\right]^{\frac{1}{2}}\frac{d\varphi}{\varphi}, 
\end{equation}
one finds from the Eq.(1) that the BD action in the Einstein frame variable  ($\tilde{g}_{\mu\nu}, \tilde{\varphi} $) , 
\begin{equation}
\tilde{{\cal A}} = \frac{1}{16 \pi } \int\sqrt{-\tilde{g}} d^{4}x \left(\tilde{R}+ 2 \tilde{g}^{\alpha\beta} \tilde{\varphi}_{,\alpha}\tilde{\varphi}_{,\beta}\right)  + \tilde{{\cal A}}_{matter}\left[ \psi_{m}, \tilde{g}_{\mu\nu}, \tilde{\varphi}\right]
\end{equation}
In the above equation derivatives are with respect to $\tilde{g}_{\mu\nu}$. Two important aspects of the Einstein frame action are that the metric and scalar field parts are untangled here, the dynamics of the gravity is governed solely by the Ricci scalar $\tilde{R}$ and secondly here matter fields couple to both $\tilde{g}_{\mu\nu}$ and $\tilde{\varphi}$. It is important to recognize that the Einstein frame energy momentum tensor is not that measured in local Lorentz frame {\it i.e.} it is not the physical energy momentum tensor. \\
The Einstein frame field equations follow by varying the action (12) with respect to $ \tilde{g}^ {\mu \nu} $ and $ \tilde{\varphi}$  
\begin{equation} 
\tilde{R}_{\alpha \beta} -\frac{1}{2}\tilde{g}_{\alpha \beta}\tilde{R}=   - 8 \pi \tilde{T}_{\alpha \beta}- 2 \left( \tilde{\varphi}_{,\alpha}\tilde{\varphi}_{,\beta}-\frac{1}{2}\tilde{g}_{\alpha\beta}\tilde{\varphi}_{,\sigma}\tilde{\varphi}^{,\sigma} \right) 
\end{equation}
\begin{equation}
\Box \tilde{\varphi} = 2\pi \frac{d ln \varphi}{d \tilde{\varphi}} \tilde{T},  
\end{equation}
The static spherically symmetric vacuum solution of the above field equations that is conformally related to the BD class I solution is the Buchdahl solution [27] which in the so called standard coordinates becomes [28] the more familiar JNW [29] or Wyman solution [30] and is given by (leaving out tilde) [47]
\begin{equation}
ds^{2}= -\left( 1-\frac{4B}{r} \right)^{\gamma} dt^{2} + \left( 1 - \frac{4B}{r}\right) ^{-\gamma} dr^{2} + \left( 1-\frac{4B}{r}\right) ^{1-\gamma} \left( r^{2} d\theta ^{2} +r^{2} sin^{2} \theta d\phi ^{2} \right)
\end{equation}
and
\begin{equation}
\varphi(r) = \sqrt{\frac{2(1-\gamma^2)}{16 \pi}} ln \left(1-\frac{4B}{r}\right), 
\end{equation}
The above form of the solution is conformally related to the BD class I solution in the standard coordinates (9). 

\section*{III. Deflection angle in the strong field regime}
Lensing theory in the strong field regime has been developed in stages by several researchers. The occurrence of relativistic images was brought forward by Darwin [9] and Atkinson [10] in their pioneering works in the field. The lens equation in the strong regime was mainly developed by Fritelli and Newman [31], Virbhadra and Ellis [12], Bozza {\it et al} [32] and Perlick [33]. After a detailed numerical study of strong field lensing produced by a Schwarzschild black hole, Virbhadra and Ellis [12] first explored observational consequences of the phenomena when the lens is the massive black hole of the galactic centre. Noting the possibility that detection of relativistic images may not be impossible in future and hence they could be used to test strong field gravity, extensive study of relativistic images started to take place. Bozza et al. [32]  developed an analytical technique of obtaining deflection angle in the strong field situation and showed that the deflection angle diverges logarithmically as light rays approach the photon sphere of a Schwarzschild black hole. Such a study was extended by Eiroa et al.[34] for lensing due to the Reissner-Nordstr\"om (RN) spacetime. Later Bozza [20] extended the method of analytical lensing for general class of static spherically symmetric metrics and demonstrated that the logarithmic divergence of deflection angle at photon sphere is a common feature for such space-times. Exploiting the Bozza's method, strong field lensing has been carried out  to several interesting cases, such as lensing due to the charged black hole of heterotic string theory [35], black holes from braneworlds [36, Einstein-Born-Infeld black holes [37], wormholes, monopole [33] etc. Very recently Bozza {\it et al} [38] have studied strong field lensing due to the Kerr black hole for equatorial observers. An interesting consequence of strong field gravitational deflection is the retro lensing [39] which occurs when the source is in between the observer and the lens or the observer is in between the source and the lens in contrast to the case of standard lensing where lens is situated in between the source the observer. The phenomena is almost same to the standard lensing except the fact that relativistic images are formed in this case for deflection angles closer to odd multiples of $\pi$ rather than even multiples. Holtz and Wheeler [39] studied retro lensing due to a Schwarzschild black hole in the Galactic bulge with the Sun as a source. Eiroa and Torres [40] considered the analytical retro lensing due to a general spherically symmetric static lens. Without remaining confined to the highly aligned case of source, lens and observer geometry, Bozza and Mancini [41] explored retro lensing due to the massive black hole of Galactic centre with the nearby (to lens) bright star S2 as source. The time delay between different relativistic images was estimated by Bozza and Mancini [42] which was later applied by several authors to some interesting cases [43]. In the present work we would employ Bozza's analytical method to obtain deflection angle in the strong field regime under the framework of the Jordan BD theory. \\
We consider the lens geometry as follows. A light ray from a source (S) is deflected by the lens (L) of mass M and reaches an observer (O). The background spacetime is taken asymptotically flat, both the source and the observer are placed in the flat spacetime. The line joining the lens and the observer (OL) is taken as the optic axis for this configuration. $\beta$ and $\theta$ are the angular position of the source and the image with respect to the optic axis respectively. The distances between observer and lens, lens and source and observer and source are $d_{ol}$, $d_{ls}$ and $d_{os}$ respectively (all distances are expressed in terms of Schwarzschild radius $r_{s}=2M$, $M$ is the mass of the lens). 
The position of the source and the image are related through the so called lens equation [12]
\begin{equation}
tan\theta -tan \beta =d \left[tan \theta +tan( \alpha -\theta) \right]
\end{equation}
where $\alpha$ is the deflection angle, $d=\frac{d_{ls}}{d_{os}}$ for standard lensing i.e. the lens is between the source and the observer and $\frac{d_{os}}{d_{ol}}$ is for retro lensing with source is in between the observer and lens. We shall skip the case of observer in between the source and the lens. For positive $\beta$, the above relation only gives images on the same side ($\theta >0$) of the source. Images on the other side can be obtained by taking negative values of $\beta$. The first and main step of getting image positions is to calculate the deflection angle.\\
For a general static and spherically symmetric spacetime of the form
\begin{equation}
ds^{2}= - A(x) dt^{2} + B(x) dx^{2} +C(x) \left(d\theta ^{2} + sin^{2} \theta d\phi ^{2} \right)
\end{equation}
where $x=r / 2M$, and as $x \rightarrow \infty$, $A(x) \rightarrow 1$, $B(x) \rightarrow 1$, $C(x) \rightarrow x^{2}$ , the deflection angle as a function of closest approach $x_{o}$ ($x_{o}=r_{o} / 2M$) is given by [48]
\begin{equation}
\alpha (x_{o})=I(x_{o})-\pi
\end{equation}
\begin{equation}
I(x_{o})= 2 \int _{x_{o}}^{\infty} \frac{ \sqrt{B(x)} dx }{ \sqrt{C(x)} \sqrt{ \frac{C(x) A(x_{o})}{ C(x_{o})A(x) }-1} } \;
\end{equation}
With the decrease of the closest approach $x_{o}$ the deflection angle will increase and for a certain value of $x_{o}$ the deflection angle will become $2\pi$ so that light ray will make a complete loop around the lens. If $x_{o}$ decreases further, light ray will wind several times around the lens before reaching the observer and finally when $x_{o}$ is equal to the radius of the photon sphere ($x_{ps}$) the deflection angle will become unboundedly large and the incident photon will be captured by the lens object. \\
Bozza develops the following technique to evaluate the integral (20) close to its divergence. The divergent integral is first splitted into two parts to separate out the divergent ($I_{D}(x_{o})$) and the regular parts ($I_{R}(x_{o})$). Then both of them are expanded  around $x_{o}=x_{ps}$ and are approximated by the leading terms. At first the integrand of Eq.(20) is expressed as a function of a new convenient variable z which is defined by 
\begin{equation}
z=\frac{A(x)-A(x_{o})}{1-A(x_{o}}
\end{equation}    
so that
\begin{equation}
I(x_{o})=\int_{0}^{1} R(z,x_{o})f(z,x_{o}) dz
\end{equation}
where
\begin{equation}
R(z, x_{o}) = \frac{2 \sqrt{A(x) B(x)}}{C(x) A^{'}(x)} \left(1-A(x_{o}) \right) \sqrt{C(x_{o})} 
\end{equation}
\begin{equation}
f(z, x_{o}) = \frac{1}{ \sqrt{ A(x_{o})-A(x) C(x_{o}) /C(x)  } }
\end{equation}
The function $R(z, x_{o})$ is regular for all values of $z$ and $ x_{o}$ but $f(z, x_{o})$ diverges as $z \rightarrow 0$ {\it i.e.} as one approaches to the photon sphere. The integral (22) is then splitted into two parts  
\begin{equation}
I(x_{o})=I_{D}(x_{o}) + I_{R}(x_{o})
\end{equation}    
where
\begin{equation}
I_{D}(x_{o})= \int_{0}^{1} R(0,x_{ps})f_{o}(z,x_{o}) dz
\end{equation}
includes the divergence and
\begin{equation}
I_{R}(x_{o})= \int_{0}^{1} g(z,x_{o}) dz
\end{equation}
is regular in z and $x_{o}$.  
The function $f_{o}(z,x_{o})$ is the expansion of the argument of the square root in the divergent function $f(z,x_{o})$ up to the second order in z
\begin{equation}
f_{o}(z,x_{o})=\frac{1}{\sqrt {p(x_{o})z+q(x_{o})z^2}}
\end{equation}    
where
\begin{equation}
p(x_{o}) = \frac{1-A(x_{o})} {C(x_{o}) A^{'}(x_{o})} \left[ C^{'} (x_{o}) A(x_{o})-C(x_{o})A^{'} (x_{o}) \right] 
\end{equation}
\begin{equation}
q(x_{o}) = \frac{ \left(1-A(x_{o})\right)^{2} }{ 2 C(x_{o}A^{'3}(x_{o}) } \left[ 2 C(x_{o}) C^{'}(x_{o})A^{' 2}(x_{o}) + \left( C(x_{o}) C^{''}(x_{o})-2 C^{'2}(x_{o}) \right) A(x_{o}) A^{'}(x_{o})-C(x_{o})C^{'}(x_{o})A(x_{o})A^{''}(x_{o}) \right]
\end{equation}
and the function $g(z,x_{o})$ is simply the difference of the original integrand and the divergent integrand    
\begin{equation}
g(z,x_{o})=R(z,x_{o})f(z,x_{o})-R(0,x_{ps})f_{o}(z,x_{o})
\end{equation}    
As $x_{o} \rightarrow x_{ps}$, $p( x_{o}) \rightarrow 0$ and hence the integral (26) diverges logarithmically. Expanding both the integral around $x_{o}=x_{ps}$ and approximating by the leading terms, Bozza obtained the analytical expression of the deflection angle close to the divergence in the form [20]
\begin{equation}
\alpha(\theta) = -u log \left( \frac{\theta D_{OL}}{b_{ps}}-1 \right) + v +O(b-b(x_{ps}))
\end{equation}
where 
\begin{equation}
u=\frac{R(0,x_{ps})}{2 \sqrt{q(x_{ps})}}
\end{equation}
\begin{equation}
v =-\pi + v_{R} +u log \frac{2 q(x_{ps})}{A(x_{ps})}
\end{equation}
\begin{equation}
v_{R}= I_{R}(x_{ps}), \; I_{R}(x_{o})= \int_{o}^{1} g(z, x_{o}) dz
\end{equation}
The coefficient $v_{R}$ may not be computed analytically for all metrics but can be evaluated numerically.
 
\section*{III A. Strong gravitational deflection due to the BD spacetime}
Since the Brans class I metric in standard coordinate is conformally related with the JNW metric, the integral $I(x_{o}$ for the metric is the same that for the JNW metric with the parameter $\gamma$ is replaced by $\frac{C+2}{2\lambda}$. But implication of this change is non-trivial. This can be easily understood from the fact that at the (first) post-Newtonian level the deflection angle for the JNW metric is $\frac{4M}{R}$, where $M=2 \gamma B$ is the gravitational mass of the lensing object and $R$ is the radius of the lensing object, which is the same that of general relativity whereas for the Jordan BD theory the deflection angle is $\frac{2M}{R} \left(1+\frac{2\omega+3}{2\omega+4}\right)$. As a result the solar system observations have so far not imposed any restriction on the parameter $\gamma$ that represents the effect of scalar field in the Einstein frame BD theory but as mentioned already the Jordan frame parameter $\omega$ is already severely constrained by the observations. \\
The radius of the photon sphere for the BD class I metric is
\begin{equation}
x_{ps}=\frac{1}{2} + \frac{2 \omega+3}{2 \omega +4}
\end{equation}
For finite $\omega$ this is smaller than the photon sphere radius of the Schwarzschild spacetime. The expression for impact parameter at photon sphere is given by  
\begin{equation}
b(x_{ps}) = \left(\frac{1}{2}+\sqrt{ \frac{2\omega+3}{2\omega+4}} \right) \left( \frac{2\sqrt{2\omega+3}+\sqrt{2\omega+4}}{2\sqrt{2\omega+3}-\sqrt{2\omega+4}} \right)^{\left(-\frac{1}{2}+\sqrt{ \frac{2\omega+3}{2\omega+4}} \right)}
\end{equation}
Exploiting the results of the strong field lensing for the JNW spacetime, the coefficients $u$ and $v$ of the deflection angle in the strong field regime for the BD class I metric have been obtained as follows 
\begin{equation}
u=1
\end{equation}
\begin{equation}
v =-\pi +  v_{R}+ log\left[ \frac{(3\omega+4)}{(2\omega+3)} \left(1-\left( \frac{2\sqrt{2\omega+3}+\sqrt{2\omega+4}}{2\sqrt{2\omega+3}-\sqrt{2\omega+4}} \right)^{\sqrt{(2\omega+3)/(2\omega+4)}} \right)^{2} \right]
\end{equation}
where 
\begin{equation}
v_{R}= 0.9496+0.1199\left(1-\sqrt{\frac{2\omega+3}{2\omega+4}} \right) + \mbox{higher order terms in} \left(1-\sqrt{\frac{2\omega+3}{2\omega+4}} \right)
\end{equation}
It can be seen that as $\omega \rightarrow \infty$, all the coefficients approach to the GR value. 

\section*{IV. Strong field observable}
Once the deflection angle is known, position of the images can be obtained from Eq.(17). In the strong field regime and when the source, lens and observer are highly aligned, the lens equation becomes [32]
\begin{equation}
\beta = \theta -d \Delta \alpha_{n}
\end{equation}
where  $ \Delta \alpha_{n}=\alpha -2n \pi $ is the offset of the deflection angle $\alpha$ and n is an integer. 
If $\theta_{n}^{0}$ are the image positions corresponding to $\alpha =2n \pi$,  the above equation gives 
\begin{equation}
\theta_{n}^{0}=\frac{b(x_{ps})}{d_{ol}}(1+e_{n})
\end{equation}
where
\begin{equation}
e_{n}=e^{(v-2n \pi)/u}
\end{equation}
and thus the position of the n-th relativistic image can be approximated as [20]
\begin{equation}
\theta_{n}=\theta_{n}^{0}+\frac{b(x_{ps})e_{n}}{u d d_{ol}} (\beta-\theta_{n}^{0})
\end{equation}
The magnification of the n-th relativistic image is given by (approximating the position of the images by $\theta_{n}^{0}$) 
\begin{equation}
\mu_{n}=\frac{1}{(\beta /\theta) \partial \beta / \partial \theta} 
\simeq e_{n}\frac{b(x_{ps})^2 (1+e_{n}) }{u \beta d d_{ol}^2}
\end{equation}
In the simplest situation if only the outermost image can be resolved as a single image then its angular separation from the remaining bunch of relativistic images is
\begin{equation}
s=\theta_{1}-\theta_{\infty}
\end{equation}
where $ \theta_{\infty}= b_{ps}/d_{ol}$ is the angular position of a set of relativistic images in the limit $ n \rightarrow \infty$. If $r$ denotes the ratio of the flux from the outermost relativistic image and those from the remaining relativistic images, then
\begin{equation}
r = \frac{\mu_{1}}{{{\infty \atop \sum} \atop n=2} \mu_{n}}
\end{equation}
For highly aligned source, lens and observer geometry these observable take the simple form 
\begin{equation}
s_{SL} = \theta_{\infty}e^{(v-2 \pi)/u}
\end{equation}
\begin{equation}
r_{SL} \simeq e^{2 \pi /u} + e^{v/u} -1
\end{equation}
for standard lensing and
\begin{equation}
s_{SL} = \theta_{\infty}e^{(v- \pi)/u}
\end{equation}
\begin{equation}
r_{RL} \simeq e^{2 \pi /u} + e^{(v+\pi)/u} -1
\end{equation}
for retro lensing. 
Since the deflection angle is already known, the strong lensing parameters {\it viz.} the position of the relativistic images, the angular separation between the outermost relativistic image and the remaining relativistic images and their flux ratio readily follow from Eqs. (48) - (51) for both standard and retro lensing. By measuring these parameters one should be able to identify the nature of the lensing object.  

\subsection*{IV A. Lensing by the super massive galactic centre}
To get an idea of the numerical values of scalar field effect in a strong lens observation we model the gravitational field of the super massive galactic centre of the Milky Way by the BD spacetime. The mass of the central object of our galaxy is estimated as $3.6 \times 10^{6}$ of solar mass and its distance is around $7.6$ kpc [14]. Therefore $d_{ol} \sim 2.14 \times 10^{10}$. Angular position of the relativistic images ($\theta_{\infty}$), the angular separation of the outermost relativistic image with the remaining bunch of relativistic images ($s$) and the relative magnification of the outermost relativistic image with respect to the other relativistic images  ($r$) are estimated by taking $\omega =500$ and $50000$ (the lower bounds obtained from two observations) for standard as well as retro lensing and are given in table 1 (magnification is converted to magnitudes: $r_{m}=2.5 Log r$). The same observable parameters when the lens is a Schwarzschild black hole are also given in the table 1 for comparison. It is clear from the table 1 that the observational predictions of the GR and the BD theory are almost the same within the given accuracy.   \\
\begin{table}
\caption {Estimates of the lensing observable in the BD theory for the central massive object of our galaxy} 
\vspace{.5cm}
\begin{tabular}
{|l|l|l|l|l|l|r|} \hline \hline
Observable & \multicolumn{3}{c|}{\em Standard Lensing} & \multicolumn{3}{c|}{\em Retro Lensing}  \\
\cline{2-7}
 & {\em Schwarzschild} & \multicolumn{2}{c|}{\em BD} & {\em Schwarzschild} & \multicolumn{2}{c|}{\em BD}\\
 \cline{3-4}  \cline{6-7}
 &  & $ \omega=500$ & $\omega=50000 $ &   & $ \omega=500$ & $\omega=50000 $   \\ \hline
&&&&&& \\
$\theta_{\infty}$  & 25.0417 & 25.0280 & 25.0415 & 25.0417 & 25.0280 & 25.0415  \\ 
($\mu$ arc sec)&&&&&& \\ \hline
s  & 0.031340 & 0.031325 & 0.031338  & 0.725217 & 0.724877 & 0.725213   \\ 
($\mu$ arc sec) &&&&&& \\ \hline
$r_{m}$ (magnitudes) & 6.82 & 6.82 & 6.82 & 6.85 & 6.85 & 6.85   \\ 
&&&&&& \\ \hline \hline
\end{tabular}
\end{table}

\section*{V. Discussion}
For certain values of coupling parameter $\omega$, the scalar tensor theories, which are among the best motivated alternatives to GR, agree with GR in post-Newtonian limit up to any desired accuracy and hence weak-field observations cannot rule out the scalar tensor theories in favor of general relativity. But in the strong field regime usually full features of a theory come in to play. As a result strong field predictions of different theories are expected to be divergent. With this anticipation in the present work strong field gravitational lensing is studied in the framework of the BD theory which is the simplest and most studied scalar tensor theory. \\
The strong field deflection angle is calculated for the BD space-time and different strong lensing parameters such as the angular positions of the relativistic images, angular separation between the outermost relativistic image and the rest of the images and also there relative magnification are estimated for both standard as well as retro lensing scenarios. It is found that all the parameters of strong field deflection in the BD theory reduce to GR values in the limit $\omega \rightarrow \infty$ as in the case of weak field lensing. The nature of such convergence is not identical but similar to the weak field scenario. This implies that against the hope there is no significant scalar field effect in the strong field observable lensing parameters. Here one may be tempted to say that the said observation was expected {\it a priori} because the radius of photon sphere in the BD theory (Eq.(36)) has $\omega$ dependence similar to that of the weak field observables of the theory such as the post-Newtonian (PPN) deflection angle [44]. It is to be noted that the radius of horizon of the BD and the Schwarzschild spacetimes are exactly the same yet the curvature components (or invariants) are very dissimilar at horizon; for Schwarzschild spacetime they are finite whereas for the BD spacetime they diverge as the horizon approaches. Hence {\it a priori} it was not possible to guess the outcome of the problem, particularly in view of the fact that the Einstein frame representation of the theory gives large deviation of strong field deflection angle parameters from those for Schwarzschild spacetime. \\
It has been already realized that observation of relativistic images is not easy [12] though see [20, 37, 38]. To observe relativistic images the resolution of the detecting telescope need to be of the order of $\mu$-arcsec or even better (the resolution achieved so far is only of the order of m-arcsec or slightly better) (whereas weak field gravitational deflection can be detected with just arcsec observational accuracy) . Proposed optical interferometer based telescopes on the International Space Station are expected to achieve angular resolution of about $0.01 \; \mu$ arcsec [45]. Hence  numerical values of the lensing parameters have been estimated at the level of nano-arcsec expressing the gravitational field due to massive compact object at the centre of the galaxy by the BD space-time. When compared with the corresponding lensing observable due to the Schwarzschild black hole, it is clear that detection of relativistic images will not give any special advantage over weak field observations to discriminate scalar tensor theories from GR. For instance observational accuracy of $0.01 \; \mu$ arcsec could yield only a bound of $\omega > 1000$ whereas with the observational accuracy of $0.1$ nano-arcsec, the lower bound of $\omega$ could be raised up to about $1.5 \times 10^{5}$. In contrast measurements of gravitational deflection of light by the solar gravity with angular precesion of $0.01 \; \mu$ arcsec could yield a bound of $\omega > 10^{8}$ [8, 44, 45]. However, strong field observations have their own merits; observation of relativistic images with finite $\omega$ would be a test for the ST gravity in the strong field regime. \\
Another interesting observation is that the strong field deflection angle in the BD theory is smaller than that of GR. Here one may recall that from weak field analysis Bekenstein and Sanders provided the theorem that {\it in a generic ST theory of gravity, the scalar field cannot enhance lensing} [46]. The present work indicates that  Bekenstein-Sanders theorem may be valid also in the strong field regime. \\

\section*{ Acknowledgment}
The authors would like to thank two anonymous referees for encouraging remarks and useful suggestions. A.B. also like to thank Dr. V. Bozza for useful and stimulating commments. \\

\end{document}